\documentclass{aa}  
\usepackage{graphicx}
\usepackage{txfonts}
\usepackage{natbib}
\begin{document}
   \title{New extended atomic data in cool star model atmospheres}

   \subtitle{Using Kurucz's new iron data in MAFAGS-OS models}

   \author{F. Grupp
          \inst{1,2}
          \and
          R.L. Kurucz \inst{3}
          \and
          K. Tan\inst{4}
          }

   \offprints{F. Grupp}

   \institute{Max Planck Institut f\"{u}r extraterrestrische Physik,
             Giessenbachstra{\ss}e, 85748 Garching, Germany\\ 
             \email{frank@grupp-astro.de}\\
         \and
              Universit\"{a}ts Sternwarte M\"{u}nchen, Scheinerstr. 1,
              81679 M\"{u}nchen, Germany\\	
              \email{frank@grupp-astro.de}\\
         \and
              Harvard-Smithsonian Center for Astrophysics, Cambridge, MA 02138, USA\\
              \email{rkurucz@cfa.harvard.edu}\\
         \and
              National Astronomical Observatories, Chinese Academy of Sciences, Beijing\,100012, P.\,R.\,China\\
              \email{tan@bao.ac.cn}
             }

   \date{Received September 15, 1996; accepted March 16, 1997}

 
  \abstract
   {Cool star model atmospheres are a common tool for the investigation
    of stellar masses, ages and elemental abundance composition. Theoretical
    atmospheric models strongly depend on the atomic data used when 
    calculating them.}
   {We present the changes in flux and temperature stratification 
   when changing from iron data computed by R.L. Kurucz in the mid 90s
   to the Kurucz 2009 iron computations.}
   {MAFAGS-OS opacity sampling atmospheres
    were recomputed with Kurucz 2009 iron atomic data as implemented in
    the VALD database by Ryabchikova.
    Temperature stratification and emergent flux distribution 
    of the new version, called MAFAGS-OS9, is compared to the former
    version and to solar flux measurements.}
   {Using the Kurucz line lists converted into the VALD 
    format and new bound-free opacities for Mg\,{\scshape i} and Al\,{\scshape i} leads to
	changes in the solar temperature stratification by not more than 28\,K.
	At the same time, the calculated solar flux distribution shows
    significantly better agreement between observations and theoretical solar models. 

    These changes in the temperature stratification of the corresponding 
    models are small, but nevertheless of a magnitude that affects stellar 
    parameter determinations and abundance analysis.}
   {}


   \maketitle
%

\section{Introduction}
   \emph{Theoretical model atmospheres} are the most basic tool
   in the investigation of stellar parameters such as temperature,
   mass, age and of the element abundances present in stellar atmospheres.
   Thus the accuracy in determining all these stellar properties strongly
   depends on the accuracy and reliability of the underlying stellar
   atmosphere.

   Many studies based on model atmospheres have shown the usefulness
   of model atmospheres for stellar parameter and abundance work. 
   The work of Fuhrmann and colleagues 
   (\citet{fuhrmann98},
   \citet{fuhrmann99}, \citet{fuhrmann02}, \citet{fuhrmann04} and
   \citet{fuhrmann08}) on nearby stars of the galactic disk and halo 
   might serve as an example of the validity of this approach. 
   While Fuhrmann stays with the assumption of local thermal equilibrium
   (LTE), many precision studies of element abundances in
   cool stars remove this simplification and calculate element
   abundances based on full radiative and collisional equilibria (Non-LTE).
   For examples, see e.g. \citet{mashonkina08a}, 
   \citet{mashonkina08b}, \citet{shi08} and \citet{mashonkina09}.

   For basic work on stellar parameters, that also
   enters the abundance analysis via the selection of the proper model,
   \citet{megessier98} , \citet{castelli94}, \citet{castelli99}, 
   \citet{grupp04b} and \citet{grupp08} show the influence of different
   model atmospheres on various stellar parameter determination methods.
   
   As a basic ingredient, bound-bound and bound-free atomic processes 
   dominate the absorption properties of late type stellar atmospheres
   over a wide spectral range.
   Taking into account as complete as possible datasets for these major
   opacity sources is shown to be vital for accurate stellar parameter and
   abundance analysis work.

\section{The model atmosphere code}
   Our study is based on the MAFAGS-OS theoretical model atmosphere 
   code described by \citet{grupp04a}. Based on the ODF version
   of T.\,Gehren in the reprogrammed version of \citet{reile87}, this 
   code relies on the following basic assumptions:
   \begin{itemize}
      \item Plane-parallel 1D geometry.
      \item Chemical homogeneity throughout the atmosphere.
      \item Hydrostatic equilibrium.
      \item Convection is treated according to the formalism of
            \citet{canuto91} and \citet{canuto92}.
      \item No chromosphere or corona.
      \item Local thermal equilibrium.
      \item Flux conservation throughout all 80 layers.
   \end{itemize}
   While these assumptions might break down for hot stars,
   stars with very extended atmospheres and the coolest stars,
   they are valid for stars in the range of T$_{eff} \approx 4000\cdots15000$\,K
   and for gravities from the main sequence down to $\log(g)\approx 0$. 
  

\section{Atomic data}
   The main target of this study is to investigate the effect of the new Kurucz 
   iron line data on models and energy distributions. These data were
   compiled for us in a private data extraction run using the VALD \citep{Kupka99} 
   database in January 2009 by Tanya Ryabchikova. This private extraction run was 
   performed including predicted levels in all elements of the iron group.

   \subsection{Old and new line list}
      \begin{figure}[ht]
        \includegraphics[width=9.0cm]{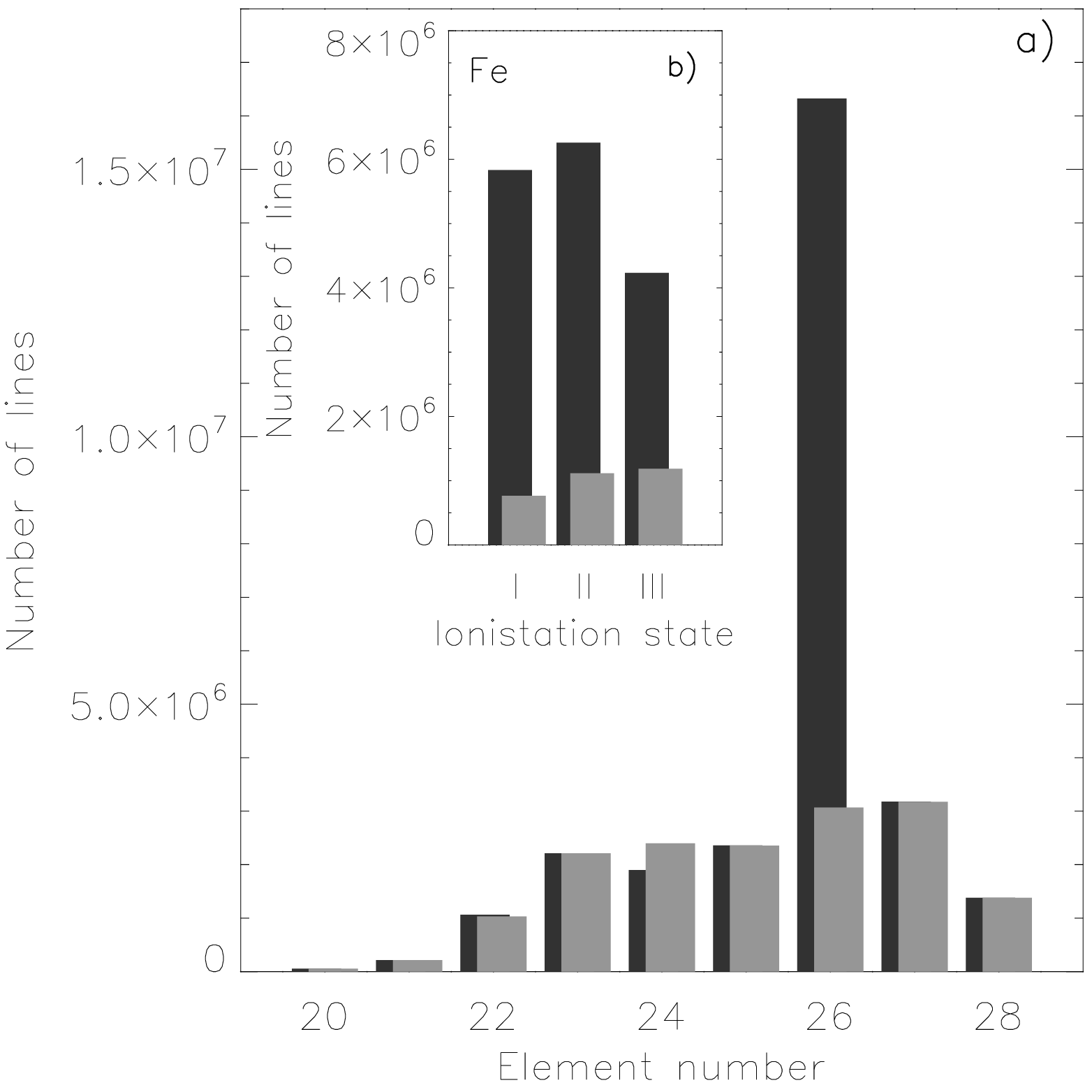}
        \caption{Number of lines for iron group elements in new\,(black) and
          old\,(grey) MAFAGS-OS line lists.\newline
                 a) For the elements Ca through Ni, sum of ionization states 
                 I,II and III.\newline
                 b) For the first three ionization states of Fe.}
        \label{linestat}
      \end{figure}   
      Our new line list was compiled based on the abovementioned data extraction
      of VALD and includes the latest Fe\,{\scshape i} to {\scshape iii} data of 
      \citet{kuruczweb}, taking into account observed and predicted transitions 
      for each ion.
      These new calculations for iron included additional laboratory levels and more
      configurations than earlier work. Thus, the number of lines connecting
      the levels greatly increased. The line list are completed for the rest of
      the iron group using  observed {\bf and} predicted levels of \citet{cd20}, 
      \citet{cd21} and \citet{cd22}. Both the new data of iron {\it and} the
      data for the rest of the iron group elements from the stated Kurucz CD-roms
      is incorporated in the VALD database used for data extraction by Tanya Ryabchikova.   
      All lines were processed using the data extraction algorithms of VALD\citep{Kupka99}.
 
      These data replace older data compiled in 1998 from \citet{cd20}, \citet{cd21},
      and \citet{cd22}. For the old line lists lines between both observed and
      predicted levels have been extract making use of the files
      GFxxyy.GAM and GFxxyy.LIN where xx stands for
      the atomic numbers 20 through 28 and yy codes the ionisation
      states 00, 01 and 02. We used our own reduction 
      software adopted from the software delivered with the CDs to process these files.  
          
      For the non iron group elements, \citet{cd23} data were used in the old
      line list. \citet{cd23} data was also included in the new list, but also
      its continuous updates incorporated in VALD. For detailed references to the 
      new list see \citet{Kupka99} and the compilations on the VALD Internet 
      platform\footnote{http://vald.astro.univie.ac.at/}. 
      Both lists - old and new - contain diatomic molecules taken from \citet{cd15} and 
      \citet{cd24}.
      
      Figure\,\ref{linestat}\,a) shows the number of lines in the final list
      for the iron group elements. The enormous increase of iron lines is 
      obvious. Small deviations in the number of Cr lines are due to the 
      fact that in the old list, Cr lines had been included both below 500\AA and 
      above 100000\AA. 
      These regions are unimportant for the solar line blending and the bug has been
      corrected in the new version of the line list.

      \begin{figure}[ht]
        \includegraphics[width=9.0cm]{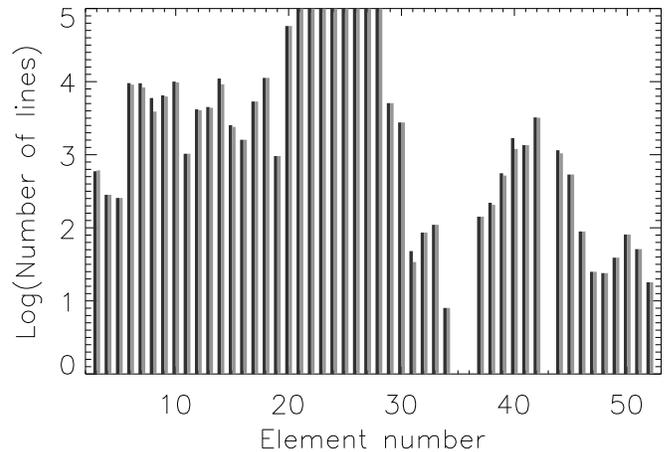}
        \caption{Number of lines for the elements $2\cdots52$ on logarithmic scale. Color
                 coding is the same as in fig.\,\ref{linestat}.}
        \label{more}
      \end{figure}
      The number of lines for the non iron group elements Li through Te is shown in
      fig.\,\ref{more}. It becomes obvious that there are few changes for all elements
      but iron between both lists.

    \subsection{Changes in continuous opacities}  
       Another change of atomic data with respect to the 2004 model is the
       inclusion of TOPBASE \citep{topbase92} bound-free cross sections for
       Al\,{\scshape i} \citep{mendoza95} and Mg\,{\scshape i} \citep{butler90}. 
       Old and new models make use of \citet{BAUTISTA97} ionization cross 
       sections for Fe\,{\scshape i} (See fig.\,\ref{AlMg}).

    \subsection{Line selection}
       Our model considers the ionization states one, two and three and runs
       with $\approx 86000$ wavelength points between 500\AA~and 850000\AA.
       Table\,\ref{selected} shows the number of bound-bound transitions
       in the new and old line-list together with the number of lines that
       were selected by the code to sample our standard solar model opacities
       \citep[For a description of the selection criteria see][]{grupp04a}. 
       The number of selected lines corresponds to our
       standard solar model. Both line lists contain additional
       4247857 lines related to transitions in diatomic
       molecules.
       \begin{table}[h]
         \caption{Data for the new and old atomic linelist.
          }
         \label{selected}
         \centering
         \begin{tabular}{c c c}
           \hline\hline
           List & Total number    & Number of selected  \\
           ~    & of atomic lines & lines for sampling \\
           \hline
           1998 & 16012015 & 322040\\
           2009 & 35352145 & 621406\\
           \hline
         \end{tabular}
       \end{table}


\section{Solar flux and temperature structure}
\begin{figure*}[ht]
      \centering	
      \includegraphics[ width=16cm]{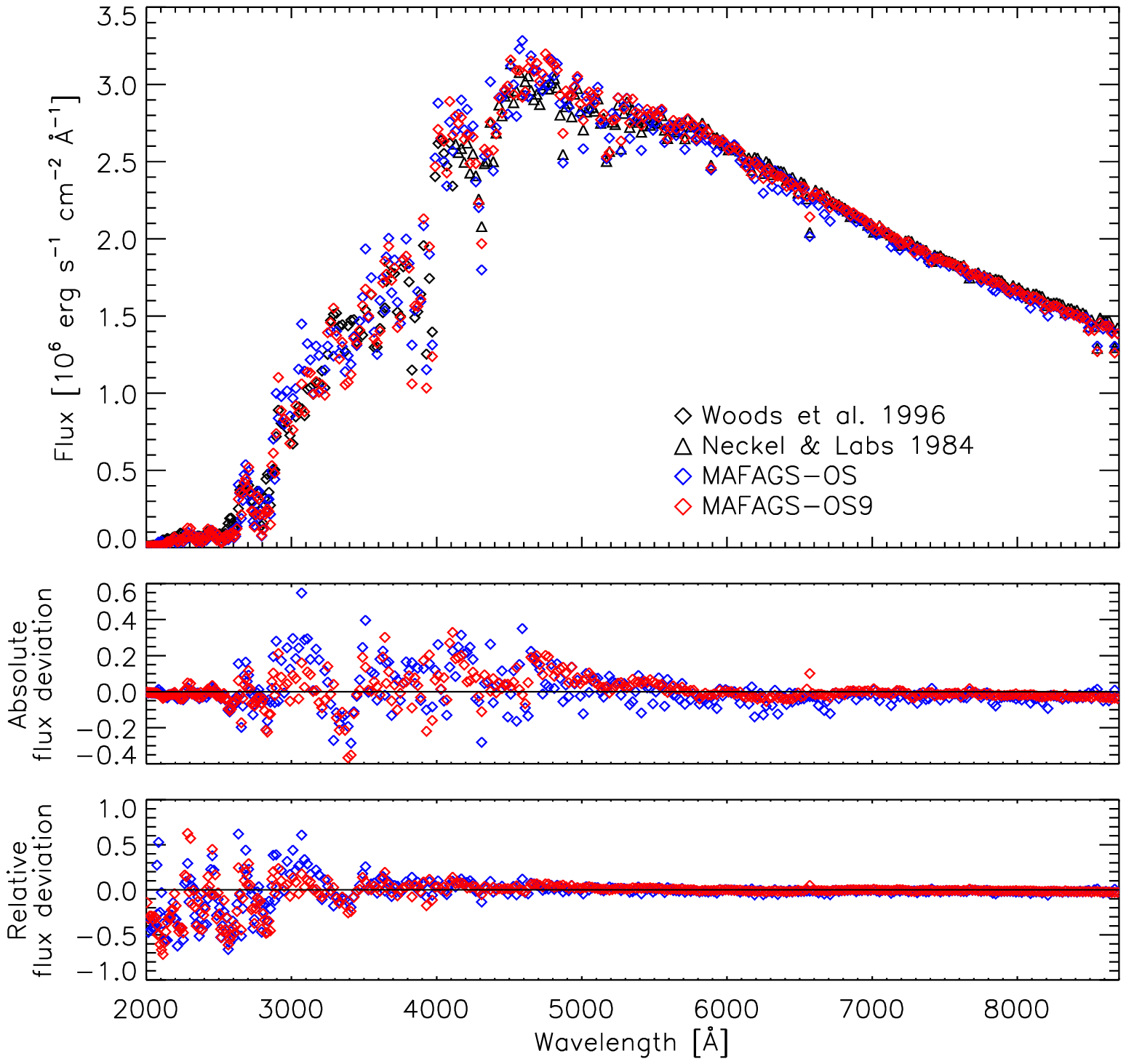}
      \caption{Comparison of observed\,(black) and model solar flux for the
               old\,(blue) and new\,(red) solar models.
               Observations are taken from \citet{neckel84} for the
               spectral range above 3300\AA and from \citet{neckel84} for the near UV
               region below 3300\AA.\newline
               Top panel: Direct comparison of solar flux data.\newline
               Mid panel: Deviation of the models from the measured solar flux in
               absolute values.\newline
               Bottom panel: Deviation of the models from the measured solar flux in
               relative values.
               }
      \label{sun_os_os9}
    \end{figure*}

    Figure\,\ref{sun_os_os9} shows the comparison of the old and new
    models with observed solar flux data from \citet{neckel84} and
    \citet{woods96}.
    Despite remaining inconsistencies in the wavelength range blueward of
    5000\AA, it is obvious that the model computed using the new iron data
    shows both less scatter and better absolute agreement with
    observed quantities.

    Nevertheless, the region between 3000\AA \ and 5000\AA \  shows remaining
    shortcomings of about $\pm 10$\% in flux, that reach $\pm 20$\% at
    some points. These discrepancies and the slight underestimation of
    the solar flux in the very scatter-dominated region below 3000\AA
    \ are present in both datasets. Still, the computations with the new 
    iron data are closer to observation even in those regions.

\begin{figure}[h]
      \includegraphics[viewport = -58 -20 340 530, width=8.cm]{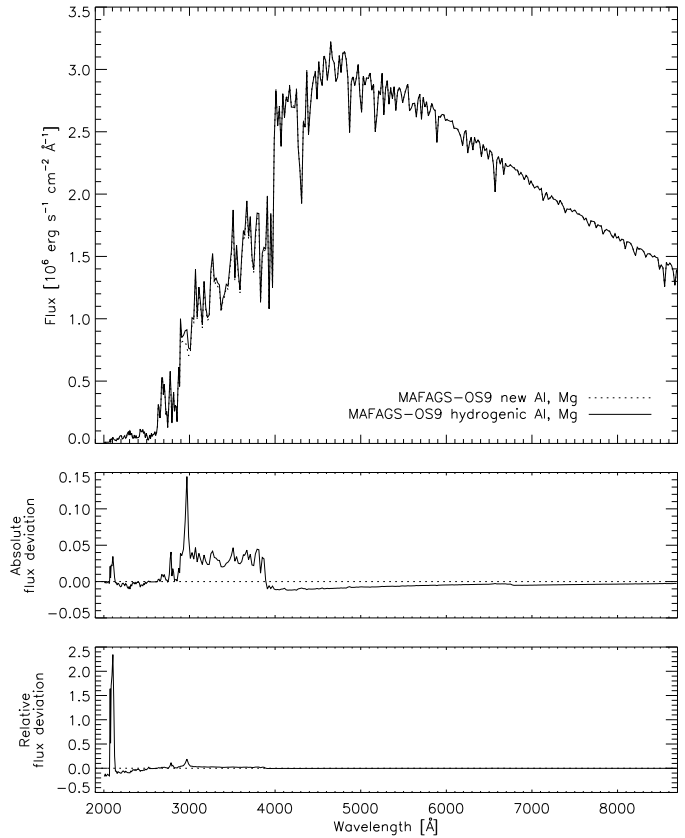}
      \caption{Comparison of MAFAGS-OS model atmospheres using ''old''
      bound free data for Al\,{\scshape i} and Mg\,{\scshape i} and recent
      detailed calculations of \citet{mendoza95} and \citet{butler90}.}
      \label{AlMg}
    \end{figure}

    The flux changes are mainly due to the new data for iron. The influence of
    new Al\,{\scshape i} \citep{mendoza95} and Mg\,{\scshape i} \citep{butler90}
    bound-free cross sections is shown in Fig.\,\ref{AlMg}. The flux differences are small in
    the visible wavelength range and contribute only to some larger deviations
    in the near ultraviolet region. These changes in the UV do not significantly 
    affect the atmosphere stratification, but they can be important for non-LTE
    calculations, as the departure coefficients of levels with transitions in the
    UV might well depend strongly on photon processes in this region 
    (see e.g. \cite{shi08}).
    The rest of the input atomic data concerning bound-free transitions is
    unchanged with respect to the ''old'' model.

    \begin{figure}[ht]
      \includegraphics[viewport = -10 -0 400 565, width=8.cm]{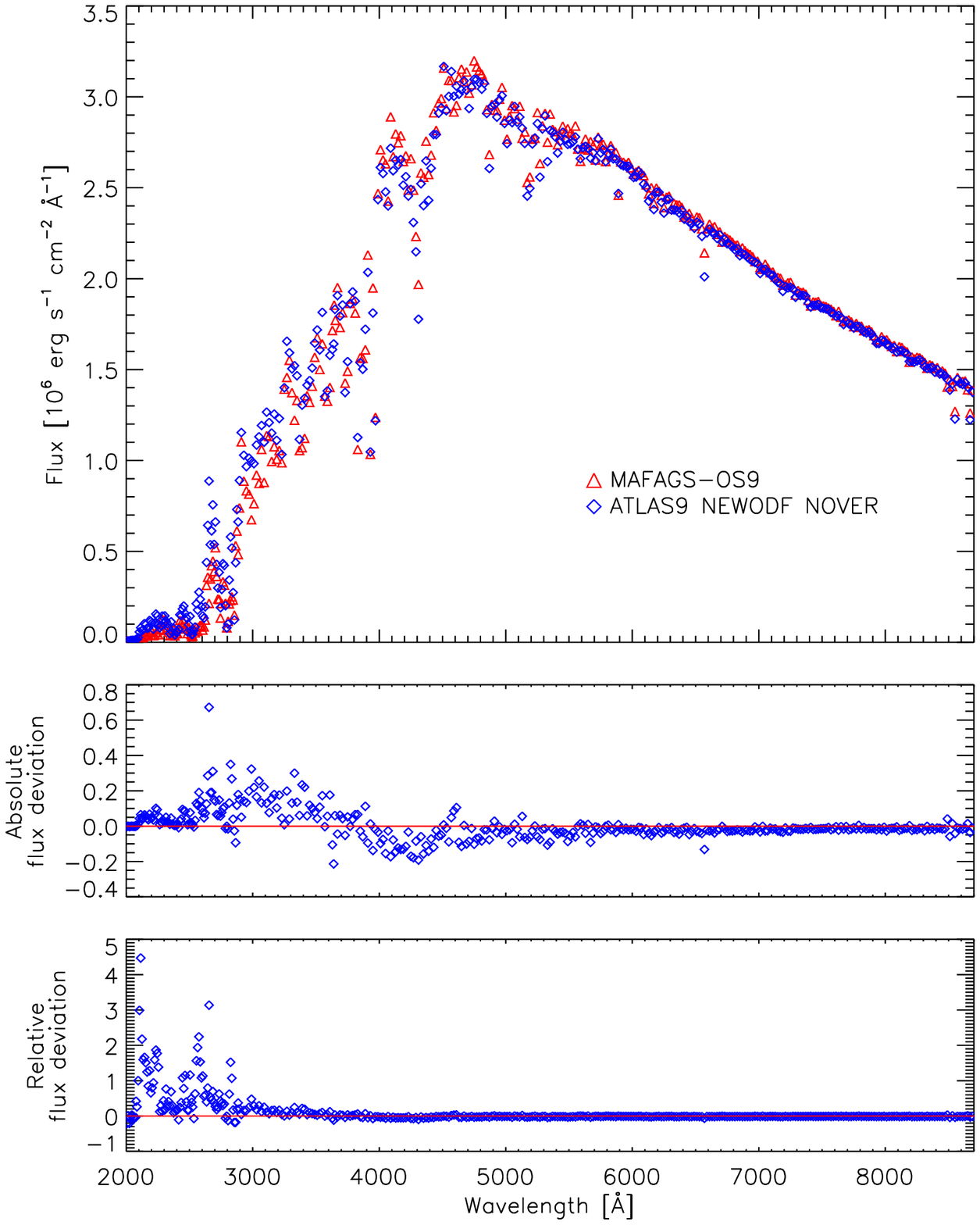}
      \caption{Comparison of the new MAFAGS-OS model\,(red) solar atmosphere with
               the recent ATLAS model with NewODF and the overshooting option
               switched off\,(blue).\newline
               Top panel: Direct comparison of solar flux data.\newline
               Mid panel: Deviation of the ATLAS model from the MAFAGS-OS9 model in
               absolute values.\newline
               Bottom panel: Deviation of the ATLAS model from the MAFAGS-OS9 model in
               relative values.}
      \label{atlas}
    \end{figure}  
    Figure\,\ref{atlas} shows the comparison of our latest model to the 
    ATLAS9 model of \citet{castelli03}. The selected ATLAS model is the 
    ASUNODFNEW  model\footnote{http://wwwuser.oat.ts.astro.it/castelli/grids.html} 
    computed with new ODFs and the overshooting option turned off.

    The ATLAS model chosen uses the new ODF
    tables of Kurucz and has the overshooting option turned off. While both models 
    show similar flux structure in the red and near IR region, with the
    ATLAS model having slightly less flux here, they significantly differ
    in the UV to green part. This leads to the ATLAS model underestimating the
    observed solar flux in the red and IR by $\approx 5$\% at 9000\AA\,
    and to a good agreement between ATLAS and
    measured solar flux in the green spectral range.   
    Looking at the picture as a whole,
    neither of the two versions fits the solar flux significantly better than the
    other, but both have good and not so good spectral regions.
    The difference between the models is quite likely to emerge from
    the lack of the new iron lines in the ATLAS9-NEWODFs and some different 
    UV continuous opacity sources used in both models. In addition, the models use
    different formalisms to treat convection, which will also influence the 
    overall flux distribution.

\begin{figure}[ht]
      \includegraphics[viewport = -53 -20 340 455, width=8.cm]{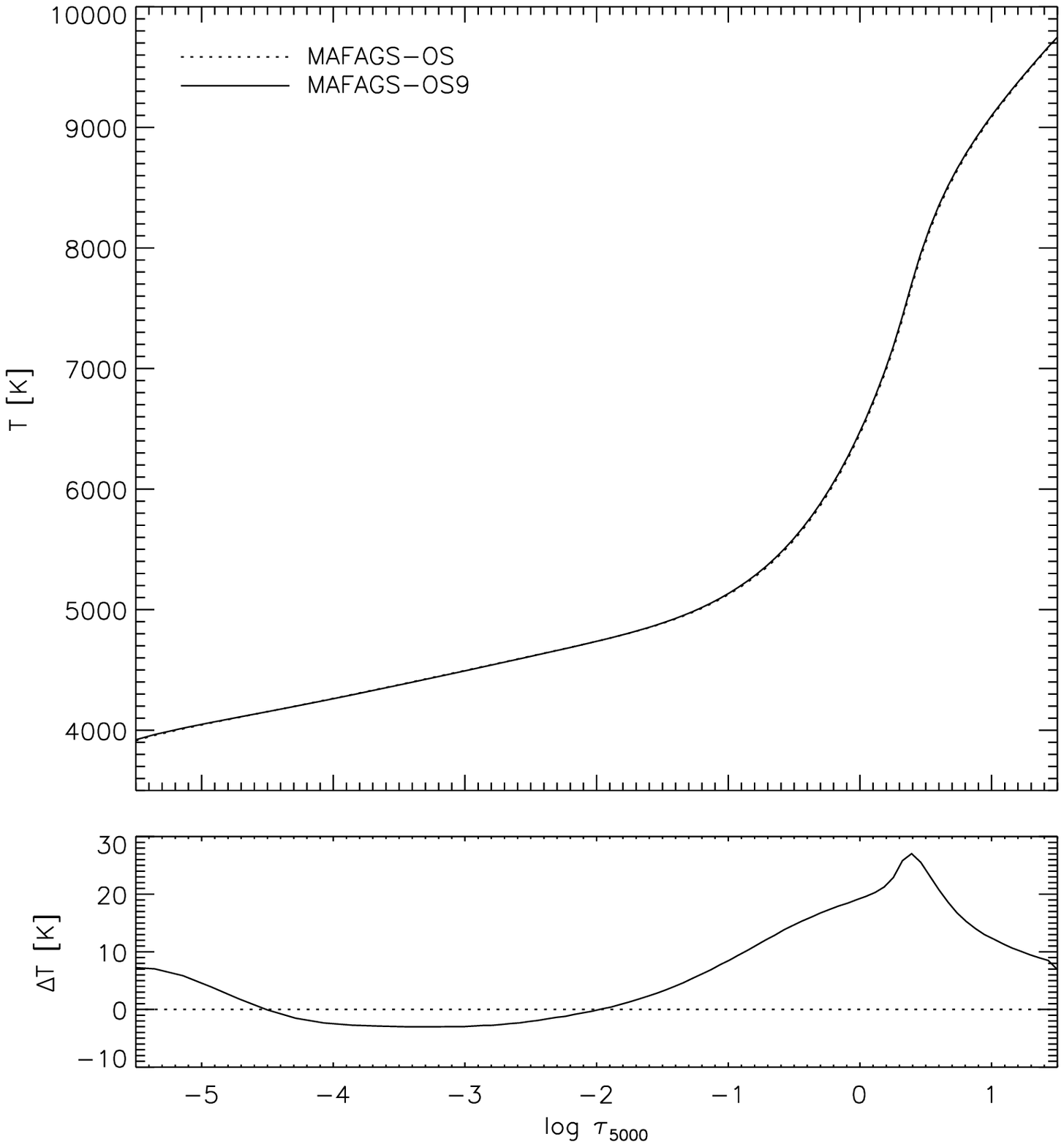}
      \caption{Comparison of the temperature structures of the old\,(dashed line)
      		   and the new\,(full line) model.}
      \label{temp}
    \end{figure}

    The changes in solar temperature structure due to the two different line lists 
    are shown in Fig.\,\ref{temp}.
    The overall effect is small, nevertheless changes of $\approx 5\cdots25$\,K 
    in the line forming regions will slightly affect line formation.
    

\section{Discussion and conclusions}
    We have shown the influence of the new iron Kurucz data on the solar
    model and solar energy distribution.
    This new data shows significantly  better agreement 
    with the observed solar flux with respect to scatter and absolute values.
    Changing to the new data also slightly changes the solar temperature 
    stratification.
    Therefore, the new data are shown to be a real improvement of the model 
    and helps to better represent solar atmospheric observables.

    Remaining discrepancies are mainly found to be in the short wavelength part
    below the solar flux maximum. They might be related to missing or inaccurate
    data in the resonances of strong bound-free absorbers or slight misplacements
    of the center wavelengths of strong lines.

    Some major implications of the improved representation of the stellar flux 
    are:\vspace{-1ex}
   \begin{itemize}
      \item Changes in synthetic colors and line indices.
      \item Changes in abundance analysis due to changes in temperature
            stratification of the model atmosphere.
      \item Changes in non-LTE analysis due to changes in the UV fluxes
            that determine photo-ionization processes emerging from 
            transitions in this spectral region.
      \item Changes in stellar parameters if they rely on synthetic 
            colors, line indices or individual lines.
      \item The lower scatter also allows better automatic comparison
            of synthetic and observed spectral data. A work field
            that shows
            increasing importance with large survey projects such as 
            SDSS \citep{york00}, LAMOST \citep{su98} and 
            GAIA \citep{perryman01} on the way. For those projects the
            mere number of spectra acquired does not allow for manual 
            inspection, requiring automated procedures of spectral analysis.  
   \end{itemize}
	Although the impact on
	the temperature stratification is relatively small, changing to
	the new data significantly changes the flux of the solar model,
	placing it into better agreement with observations.

\begin{acknowledgements}
      The authors would like to thank Lyudmila Mashonkina for her 
      help in testing and improving MAFAGS-OS in the UV spectral
      region. Additional thanks go to Thomas Gehren for his
      continuing support and for sharing his enormous experience
      with us. We also thank Zhao Gang for promoting our
      Sino-German collaboration.\\
      Our special thanks to Tanya Ryabchikova, Nikolai Piskunov
      and Denis Shulyak who, by providing the VALD compiled data, made 
      this work possible.\\
      Additional thanks  to the referee Fiorella Castelli, through
      her comments a wrong setting in the line selection was found and
      corrected.\\
      Part of this work was supported by the German
      \emph{Deut\-sche For\-schungs\-ge\-mein\-schaft, DFG\/} project
      number Ge~490/31--1, by the National Natural Science 
      Foundation of China under Grants No. 10521001 and 10433010 and by the 
      National Basic Research Program of China (973 Program) under 
      Grant No. 2007CB815103.\newpage
\end{acknowledgements}

\bibliography{./astro}
\bibliographystyle{aa}





\end{document}